\documentclass[iop,apj,numberedappendix,twocolappendix,appendixfloats]{emulateapj}
\usepackage{epsfig}
\usepackage{amssymb}
\usepackage{amsmath}
\usepackage{natbib}

\newcommand{\Msun}{{M_\odot}}
\newcommand{\Mbh}{M_{\rm BH}}

\newcommand{\Ledd}{L_{\rm Edd}}

\def \cm{~\rm{cm}}

\def \K{~\rm{K}}
\def \g{~\rm{g}}

\shorttitle{On the Virialization of Disk Winds}
\shortauthors{Kashi et al.}

\begin{document}

\title{On the Virialization of Disk Winds: Implications for the Black Hole Mass Estimates in AGN}

\author{
Amit Kashi\altaffilmark{1},
Daniel Proga\altaffilmark{1},
Kentaro Nagamine\altaffilmark{1,2},
Jenny Greene\altaffilmark{3}, and 
Aaron J. Barth\altaffilmark{4}
}

\altaffiltext{1}{Department of Physics \& Astronomy, University of Nevada, Las Vegas, 4505 S. Maryland Pkwy, Las Vegas, NV, 89154-4002, USA ~~  \href{mailto:kashia@physics.unlv.edu}{kashia@physics.unlv.edu}}
\altaffiltext{2}{Department of Earth and Space Science, Graduate School of Science, Osaka University, 
1-1 Machikaneyama-cho, Toyonaka, Osaka, 560-0043, Japan}
\altaffiltext{3}{Department of Astrophysical Sciences, Princeton University, Princeton, NJ 08544, USA}
\altaffiltext{4}{Department of Physics and Astronomy, University of California, Irvine, Irvine 92697, CA, USA}

\begin{abstract}
Estimating the mass of a supermassive black hole (SMBH)
in an active galactic nucleus (AGN) usually relies on
the assumption that the broad line region (BLR) is virialized.
However, this assumption seems invalid in
BLR models that consists of an accretion disk and its wind.
The disk is likely Keplerian and therefore virialized.
However, the wind material must, beyond a certain point,
be dominated by an outward force that is stronger than gravity.
Here, we analyze hydrodynamic simulations of four different disk
winds: an isothermal wind, a thermal wind from an X-ray
heated disk, and two line-driven winds, one with and the other
without X-ray heating and cooling. For each model, we check
whether gravity governs the flow properties, by
computing and analyzing the volume-integrated quantities that appear
in the virial theorem: internal, kinetic, and gravitational energies,
We find that in the first two models,
the winds are non-virialized whereas the two line-driven disk
winds are virialized up to a relatively large distance.
The line-driven winds are virialized because they accelerate
slowly so that the rotational velocity is dominant and
the wind base is very dense. For the two virialized winds, the so-called
projected virial factor scales with inclination angle
as $1/ \sin^2{i}$. Finally, we demonstrate that an outflow from
a Keplerian disk becomes unvirialized more slowly when it conserves
the gas specific angular momentum  --  as in the models considered here,
than when it conserves the angular velocity
-- as in the so-called magneto-centrifugal winds.
\end{abstract}


\section{Introduction}
\label{sec:intro}

Many astrophysical systems in equilibrium are virialized.
Examples range from objects that are in hydrostatic equilibrium, 
such as stars, planets, and intergalactic medium, to dynamical systems
such as planetary systems, binary stars, stellar globular clusters, and galaxies.
For such systems, the mass inside a sphere of a radius $r$ and a characteristic velocity $v$ are related through the well known equation
\begin{equation}
M(<r)=f \frac{r v^2}{G},  
\label{eq:MBH}
\end{equation}
where $f$ is a factor that depends on the geometry and dynamics, and $G$ is the gravitational constant.
For example,  in the case of Keplerian rotation, $f=1$ because the gravity and centrifugal 
forces are equated, while in the case of supersonic accretion, $f=0.5$ because 
the gravitational potential energy and kinetic energy are equated. 
For such simple cases, $f$ does not change when it is integrated over some volume, even if weighted by a non-uniform density.
If the system is more complicated, the density-weighted, volume-integrated $f$ could be dominated by the denser part of the flow.

It is often assumed that the broad line regions (BLRs) in active galactic nuclei (AGN) are also virialized. This assumption cannot be directly
verified because the BLRs are spatially unresolved. However, the very compactness of the BLRs and
the broadness of the observed lines suggest that the dynamics of the BLR gas is strongly coupled to the gravity of 
the central supermassive black hole (SMBH). Therefore, the assumption of virialization has been used to determine 
the SMBH mass $\Mbh$, provided both $v$ and $r$ are known.
In the case of AGN, practically all the mass is concentrated in the center 
with $\Mbh$ being much greater than that of stars and gas, therefore $M(<r) = {\rm constant} = \Mbh$. 
The emission line width, $\Delta v$, can be measured relatively easily and used as an estimate of $v$.
Using reverberation mapping, the distance $r$ can be estimated 
via $r=c\tau$, where $\tau$ is the time delay for BLRs to respond 
to changes in the continuum (e.g., \citealt{BlandfordMcKee1982}; 
\citealt{Peterson1993}).
This method has been used in many AGN surveys for multiple emission lines with different characteristic emission radii, and it has been improved 
over the years
(e.g., \citealt{Ulrich1997}; \citealt{PetersonWandel1999}, \citeyear{PetersonWandel2000}; \citealt{Peterson2001},\citeyear{{Peterson2004}}; \citealt{Kaspi2000}; \citealt{Kollatschny2003}; \citealt{Bentz2009}; \citealt{Pancoast2011}; \citealt{HryniewiczCzerny2012}; \citealt{Shen2013}; \citealt{Barth2013}).
It has also been suggested that the BLRs are virialized based on the correlation between the time delay and line width
(e.g., \citealt{PetersonWandel1999}, \citeyear{PetersonWandel2000}; \citealt{OnkenPeterson2002};  \citealt{Kollatschny2003}; \citealt{Peterson2004}).

Although the basic assumptions behind the above mentioned method
are very plausible, the method has its limitations.
For example, different BLR structures, the radiation pressure, viewing angle, 
gravity due to the host galaxy, and different methods for characterizing 
the broad-line widths are all expected to affect the measurements 
(e.g., \citealt{Krolik2001}; \citealt{Watson2007}; \citealt{Marconi2008}; 
\citealt{Gaskell2010}; \citealt{Goad2012}; \citealt{Denney2012}).
Due to these systematic effects, the value of $f$ may be uncertain by two 
orders of magnitude (e.g., \citealt{Krolik2001}).
Observations have estimated the average values in the range of 
$\langle f \rangle \approx 1$--$6$ 
\citep{McLureDunlop2004, Onken2004, Woo2010, Grier2013}.

Observational estimates of the value of $f$ may differ from the theoretical ones.
In some cases, the value of $f$ used by observers is composed of a correction factor which depends on the definition of the line width,
and a theoretical factor which assumes some geometry and dynamics.

Another issue with this method is that the relation $r \propto v^{-2}$, which comes from the virial assumption 
(Eq.~\ref{eq:MBH}),
can hold not only in virialized systems, but also in several types of
winds \citep{Krolik2001}. On the other hand,
winds have been proposed as one of the possible scenarios for the BLR in AGNs, and have been studied extensively both observationally and theoretically (e.g., \citealt{Kallman1993}; \citealt{Arav1994}; \citealt{Konigl1994}; \citealt{Murray1995}; \citealt{ChiangMurray1996}; \citealt{Nicastro2000}; \citealt{Laor2006}; \citealt{Bentz2010}; \citealt{Sim2010}; \citealt{Wang2011}; \citealt{Roth2012}; \citealt{Kollatschny2013}).

A closely related question is how observations of a certain emission line 
can tell us whether the system is virialized or not.
\cite{Richards2011} showed that a number of different emission-line features are consistent 
with a two-component disk$+$wind model of the BLR 
(e.g., \citealt{Collin-Souffrin1988}; \citealt{Leighly2004}).
\cite{BaskinLaor2005} found that, while the most widely used H$\beta$ line is virialized, the second most used C\textsc{iv} line has a non-virialized component (but see \citealt{Denney2013} for recent contradicting results).
This may suggest that virialization can be limited to only some parts of the BLRs.

These considerations motivate us to check whether the outflows could be 
virialized and if so to what extent.
In this paper, we analyze four simulations of different {\it disk wind} models, 
and examine which parts of the winds are virialized.
In \S\,\ref{sec:Simulations}, we briefly describe the simulations that we analyze. 
In \S\,\ref{sec:Methods}, we describe our analysis methods. 
In \S\,\ref{sec:Results}, we present our analysis results. 
In \S\,\ref{sec:Discussion}, we discuss how an outflow from
a virialized Keplerian disk becomes unvirialized.
A summary of our findings is given in \S\,\ref{sec:Summary}.

\section{Simulations}
\label{sec:Simulations}

\begin{table*}[!ht]
\caption{
Comparison of physics in the simulations analyzed in this work.
References:
Model 1. \cite{GiustiniProga2012};  Model 2. \cite{Luketic2010}; Model 3. \cite{Proga2003}; Model 4. \cite{ProgaKallman2004}.
}
\begin{center}
\begin{tabular}{ccccccll}
\hline \hline
Model           & Gravity  & Rotation  & Radiative gas       & Radiation  & Equation of  & $L_{\rm{tot}}/L_{\rm{Edd}}$  & Units of          \\
                &          &           & heating and cooling & pressure   & state        &                              & distance          \\
\hline
1               & Yes      & Yes       &  No                 &  No        & Isothermal   & 0                            & $GM/c_s^2$        \\

2               & Yes      & Yes       &  Yes                &  No        & Adiabatic    & 0.03                         & $GM/c_s^2$         \\

3               & Yes      & Yes       &  No                 &  Yes       & Isothermal   & 0.0015                 & $r_{\rm{WD}}=8700~\rm{km}$ \\

4               & Yes      & Yes       &  Yes                &  Yes       & Adiabatic    & 0.6                          & $2GM/c^2$   \\

\hline
\end{tabular}
\end{center}
\label{Table:comparemodels}
\end{table*}

Here, we analyze four different axisymmetric two-dimensional (2-D) Eulerian hydrodynamic simulations of
winds driven off disks that accrete onto a central object.
These simulations are chosen from the earlier works by Proga and his
collaborators to provide controlled environments, representing
different aspects of winds driven from a Keplerian accretion disk. 
We chose these four particular simulations in order to explore
disk winds with different physics not because they
are necessarily the best to model the BLRs.
In fact, only the last simulation has been performed with parameters suitable for quasars and captures the minimum required physics.
However, the first three simulations are relevant in the context of this paper because they can be rescaled to AGN (see below for scaling relationships). In addition, as we show, they are useful to understand and highlight the properties of the wind in the fourth simulation and asses its generality.
Specifically, the first two simulations illustrate that not all disk winds appear to be virialized and emphasizes
that the size of the acceleration zone is the key to wind virialization.
Table~\ref{Table:comparemodels} compares the physics included in each model.
Below we give brief descriptions of each model. 

\vspace{2 mm}
{\bf Model 1:} 
An isothermal, steady state, wind model~B from
\citet{GiustiniProga2012}. This model is one of several tests computed by \citet{Luketic2010}.
In this, perhaps simplest, disk wind models, the gas expands and
accelerates because it is implicitly heated (gains energy) in accordance with the assumption of being isothermal. 
The calculations and results are presented in units where $GM=1$ and the sound speed $c_s=1$ (i.e., distance is in units of $r_0=GM/c_s^2$).
As found by \citet{Font2004},
the geometry of such winds depends on the density profile along the equator:
flat profiles yield vertical winds whereas the profiles
where the density decreases strongly with radius yield
spherical winds
(see also \citealt{GiustiniProga2012} for more details).
Here, we present the results for the density profile 
$\rho \propto r^{-2}$ that produces a nearly spherically symmetric wind.
The ratio between the maximum velocity at the end of the computational domain 
and the Keplerian velocity at $r=1$ (in the simulation units) 
is $v_{\rm out,max}/v_{\rm K}(r=1)=3.57$ for this model. 
This ratio is an indicator of the wind acceleration and virialization: lower
values indicate weak acceleration (the gravity is dominant)
whereas high values suggest very strong acceleration resulting
in an unvirialized outflow. A value of more than 3, as in model 1,
indicates an unvirialized outflow. This is confirmed by our more
detailed analysis summarized below in \S\,\ref{sec:Results}.

\vspace{2 mm}
{\bf Model 2:} A thermal wind model from an X-ray heated disk, which
is similar to the model C8 of \citet{Luketic2010}, but with an outer
radius 5 times greater.
The BH mass in this model is $M=7 \Msun$, and the total luminosity is 
$L_{\rm{tot}}=0.03\Ledd$, where $\Ledd$ is the Eddington luminosity 
(therefore the wind cannot be driven by radiation pressure).
The adiabatic index is $\gamma=5/3$.
The density profile is $\rho =10^{-11} (r/r_{\rm{IC}})^{-2} \g \cm^{-3}$, where $r_{\rm{IC}}= GM \mu m_p/(k_B T_{\rm{IC}}) = 4.8 \times 10^{11} ~(M/7\Msun) ~(\mu/0.6) ~(T_{\rm{IC}}/1.4 \times 10^7 \K)^{-1} \cm$ is the inverse-Compton (IC) radius.
The distance for this model is in units of $r_{\rm{IC}}$.
We note that the units for the Model 1 and 2 are  related to each other
by $r_{\rm{IC}}=r_0=GM/c_s^2$. As Model~1 is a simplest example
of a thermally driven disk wind, Model~2 can be viewed as a more
physical variate of Model~1 because in Model~2 the physics of gas
heating and cooling is explicitly included.
One of the consequences of this more physical model is that
here the wind solution does not depend on the density profile as long as the density is high and the gas temperature is low.
This model includes radiation heating and cooling, as applied in \citet{ProgaKallman2002} and \citet{Proga2000}.
The local X-ray flux is corrected for optical depth effects, taking only electron-scattering as the source of opacity.
This model produces a fast-moving (few$\times 100 ~\rm{km}~{\rm s}^{-1}$) high-density wind with convex streamlines close to the poles and slower 
(few$\times 10 ~\rm{km}~{\rm s}^{-1}$) concave streamlines close 
to the equator.
The $v_{\rm out,max}/v_{\rm K}(r=1)$ ratio is 1.50 indicating
that the acceleration is weaker compared to model 2.

\vspace{2 mm}
{\bf Model 3:} A line-driven disk wind model E2 from \citet{Proga2003} \citep[see also][]{Proga1998}. 
In this model thermal driving is negligible, the driving is not due to
gain of thermal energy but due to transfer of momentum from radiation
to the gas.  As such, compared to Model~1 and 2, Model~3 represents a physically different class
of winds. 
The model assumes a disk around a $0.6 \Msun$ white dwarf (WD) of radius
$r_{\rm{WD}}=8700~\rm{km}$ with an isothermal equation of state.
We show this particular model because it quite well accounts
for observations of disk winds in cataclysmic variables \citep[see
fig. 2 in][]{Proga2003} and has been used as a base for disk winds in
AGNs (see below).
Distances in this model are expressed in units of $r_{\rm{WD}}$.
The total luminosity of the accretion disk and WD is $L_{\rm{tot}}=1.5\times10^{-3} \Ledd$.
The model computes the radiation force due to lines using the intensity of the radiation integrated over the UV-band only.
The density profile in the disk is $\rho =10^{-9} (r/r_{\rm{WD}})^{-2} \g \cm^{-3}$.
The model calculates the line force that drives winds from a thin disk based on \cite{Proga1999}.
The resulting outflow shows radial streamlines with very high velocities (few$\times 10^3 ~\rm{km}~{\rm s}^{-1}$) in high latitudes and very low (few$\times 10 ~\rm{km}~{\rm s}^{-1}$) velocities at low latitudes.
Here, the $v_{\rm out,max}/v_{\rm K}(r=1)$ ratio is 1.99. Rescaling
results from Model~3 to AGN is not straightforward, in part because
it does not include some of the physical processes that are essential
in AGN, e.g., the X-ray ionization. Therefore, our last model
is an extension Model~3  that was computed specifically for AGN.

\vspace{2 mm}
{\bf Model 4:} A line-driven wind with X-ray heating and cooling (\citealt{ProgaKallman2004}).
The model describes a wind from a disk around a $M=10^8 M_\odot$ SMBH.
The distance in this model is given in units of the Schwarzschild radius $r_s=2GM/c^2=3 \times 10^{13} \cm $.
The disk luminosity is $L_D=0.5 \Ledd$, and the luminosity
of the central engine is $L_c=0.1 \Ledd$ with $90\%$ of the radiation in the UV and $10\%$ in the X-ray.
The model computes the radiation force due to lines using the intensity of the radiation integrated over the UV-band only. 
The central engine produces photons that can ionize the gas, but its contribution as a source of radiation pressure was excluded.
The adiabatic index is $\gamma=5/3$.
The gas density along the disk midplane was assumed to scale as $\propto r^{-2}$.
For small radii at the disk atmosphere and wind base, the model predicts a typical density
$\sim 10^{-12} \g \cm^{-3}$ which results in a relatively low
photo-ionization parameter ($\log \xi <-5$) despite the strong radiation
coming from the center.
In addition, the model predicts significant self-shielding:
dense clumps form close to the center (``failed wind'') as a result of the
over-ionization,  which provide shielding for the gas launched at large radii.
The disk wind is very fast ($\sim 10^4 ~\rm{km}~{\rm s}^{-1}$) 
at low latitudes whereas at high latitudes, there is a low density inflow.
Here, the $v_{\rm out,max}/v_{\rm K}(r=1)$ is 0.13. As shown by
\citet{ProgaKallman2004} this class of models well accounts for the
properties of outflows observed in broad absorption line
quasars \citep[see also][and references therein]{Sim2010}.

\vspace{2 mm}

\section{Analysis Methods}
\label{sec:Methods}

The simulations described above solve either hydrodynamic or radiation$+$hydrodynamic 
equations in an Eulerian form on a 2-D grid with axial symmetry. 
The wind solution is given as the spatial distribution of local quantities as a function of time.
These quantities are the density $\rho$, specific internal energy $e$, and velocity $\mathbf{v}$.
Therefore the simulations provide all the necessary information to compute the terms of the virial equation
\begin{equation}
\Phi_G=-2(E+K),
\label{eq:virialeq}
\end{equation}
where $\Phi_G$, $E$ and $K$ are the density-weighted, volume-integrated quantities of gravitational potential $\phi_G=GM/r$,  specific internal energy, and specific kinetic energy $k=v^2/2$, respectively.

For Models 1--4, we compute the terms in the virial theorem.
These models assumed axial symmetry, however, the rotational component of velocity $v_\varphi$  was implicitly calculated, and it is used in our analysis.
Following the original papers that presented the simulations, we use spherical polar coordinates.

We calculate the kinetic component of the virial factor
\begin{equation}
f_k=\frac{|\phi_G|}{2k}
\label{eq:f_k}
\end{equation}
as a function of position.
When $e \ll k$, $f_k$ measures where the flow is close to or largely deviating from virialization.
In this case  $f_k \simeq 1$ would indicate a virialized region in the flow.
We note that in some cases $e$ can be dominant, for example, in stellar interiors.

We calculate the density-weighted, surface-integrated viral quantities using the following equation:
\begin{equation}
\tilde{q}_i = \int\limits_{\varphi=0}^{2\pi}\ \int\limits_{\theta=0}^{\pi}\ q_i \rho^n \sin\theta \,d\theta\,d\varphi,
\label{eq:tildeqi}
\end{equation}
where $q_i = (\phi_G, e, k)$, and they are weighted by $\rho^n$.  
To examine the effects of winds on observations, we take $n=1$ for continuum fluorescence excitation line emission, and $n=2$ for recombination line emission and collisionally excited line emission.

Finally, we compute the density-weighted, volume-integrated quantities
\begin{equation}
Q_i = \int\limits_{\varphi=0}^{2\pi}\ \int\limits_{\theta=0}^{\pi}\ \int\limits_{r=0}^r \ q_i \rho r^2 \sin\theta \,dr\,d\theta\,d\varphi,
\label{Qi}
\end{equation}
where $Q_i = (\Phi_G, E, K)$ for $q_i = (\phi_G, e, k)$, respectively.  
Also, $K = (K_r, K_\theta, K_\varphi$) are the radial, meridional, and rotational components of $K$, respectively. 

We use the local properties of the wind to check if $k$ scales with radius the same way as $\phi_G$, namely, if $f_k$ is radius independent.
If $f_k$ is constant, then it means that Equation~(\ref{eq:virialeq}) will hold and the system is virialized.
But even if it is not constant, Eq.~(\ref{eq:virialeq}) can still hold when the density-weighted volume integral is performed, and the system is virialized.

In Appendix A, we illustrate the results of our analysis on the well known Bondi accretion flow \citep{Bondi1952} and Parker wind \citep{Parker1965}.
We will use these results to compare with more complicated cases in the following sections.

\section{Results}
\label{sec:Results}

\begin{figure*}
\textbf{\Large (a)}
\newline
\includegraphics[trim=0cm 0.17cm 0cm 0.65cm, clip=true, width=1.08\textwidth]{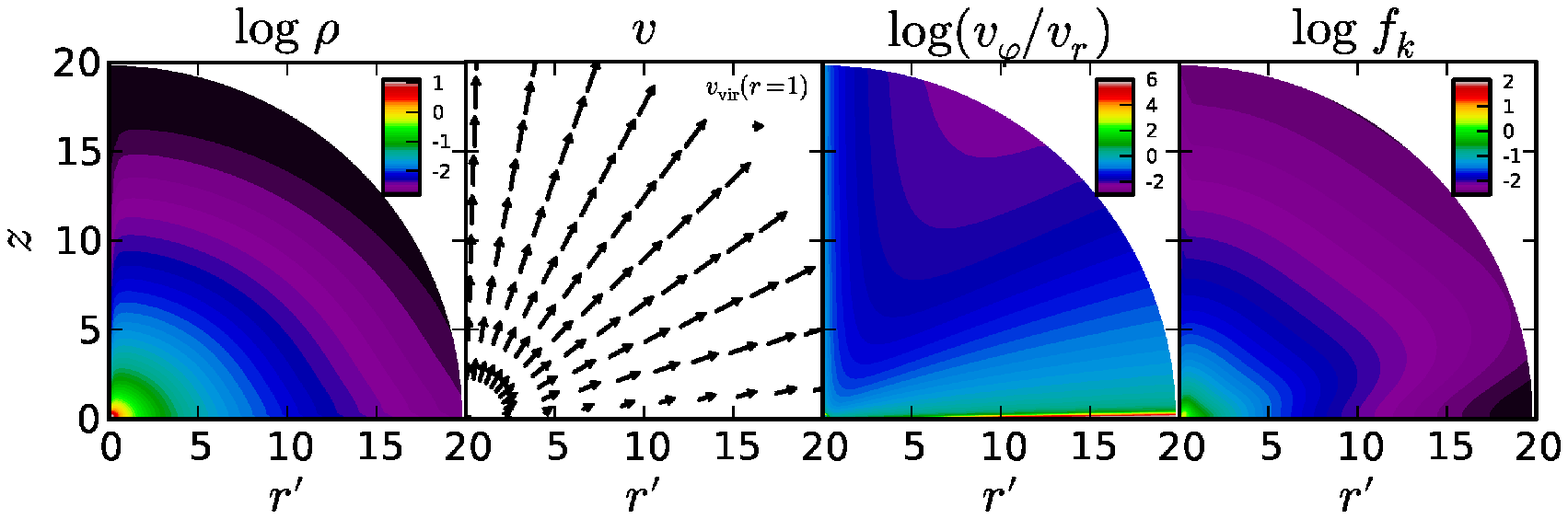}
\textbf{\Large (b)}
\newline
\includegraphics[trim=0cm 0.5cm 0cm 0.5cm, clip=true, width=1\textwidth]{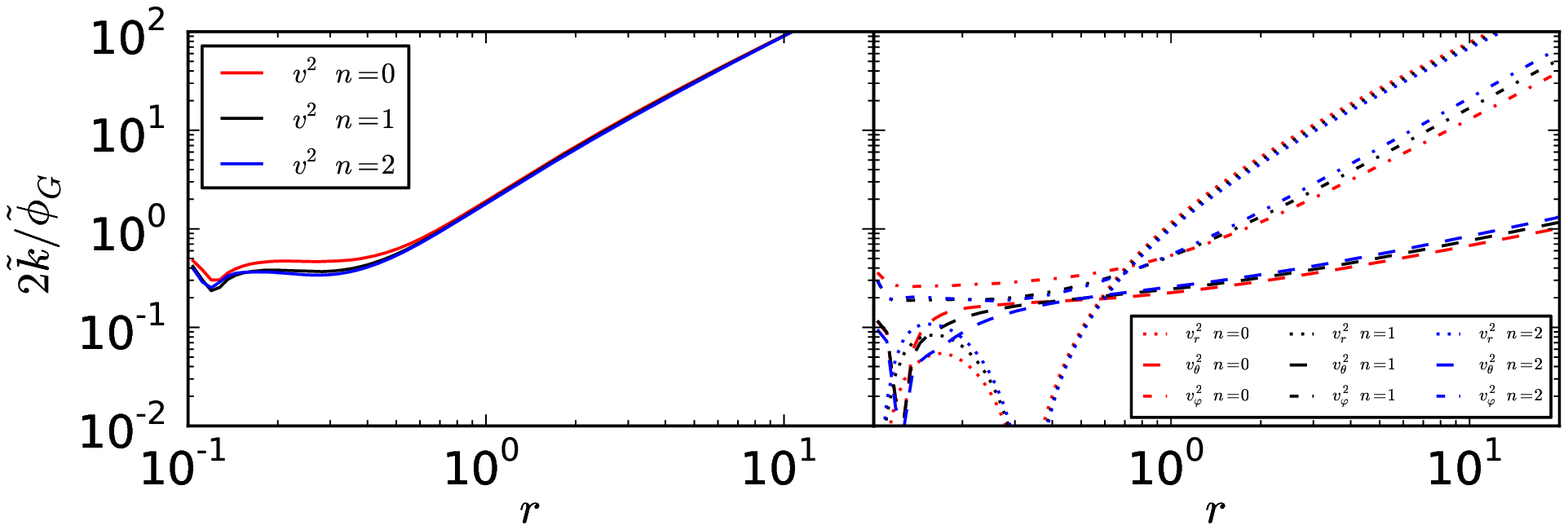}\\
\textbf{\Large (c)}
\newline
\includegraphics[trim=0cm 0.6cm 0cm 0.55cm, clip=true, width=1\textwidth]{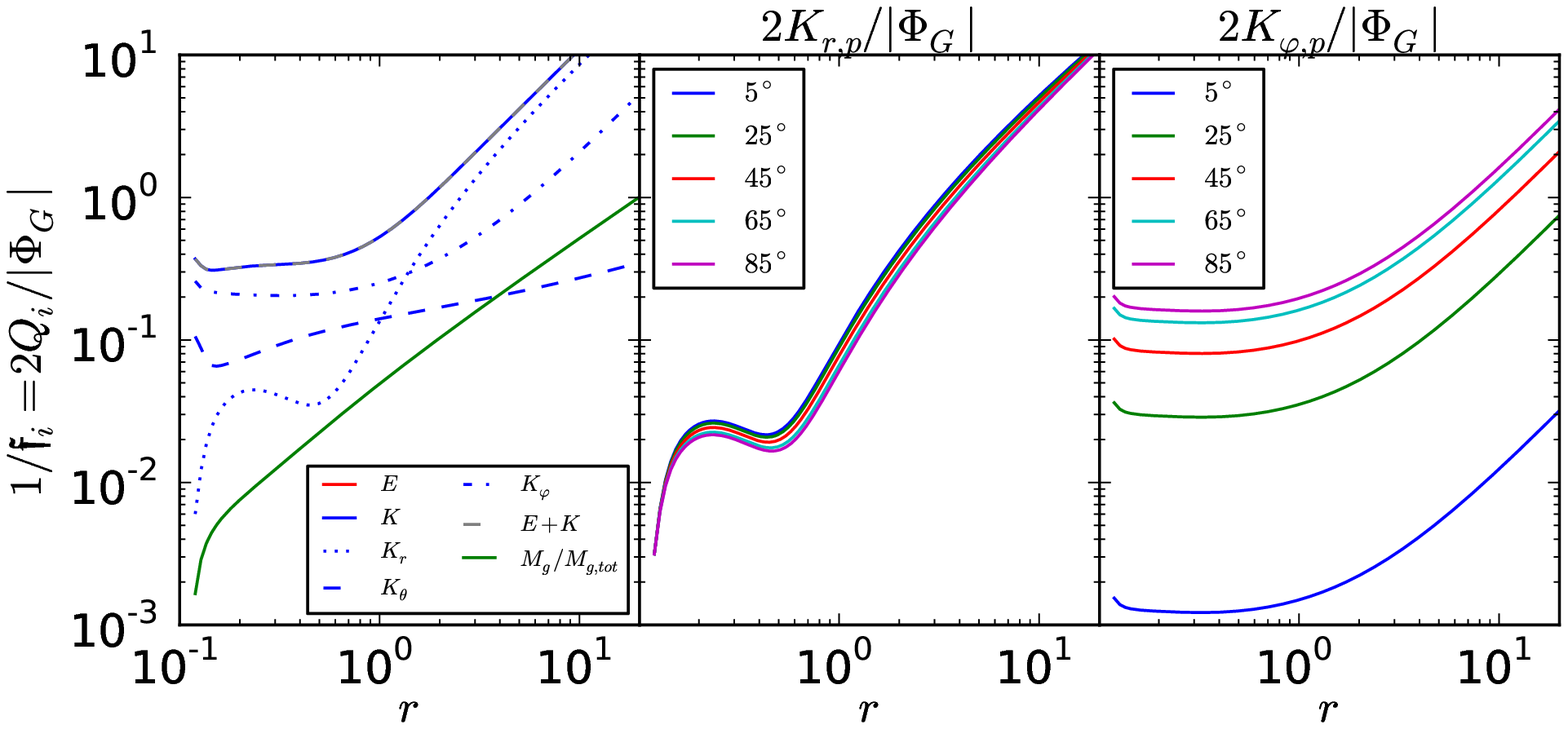}
\caption{{\small
For Model 1 (isothermal wind).
\textbf{Subfigure (a):} The flow properties are computed and presented in units where $GM=1$.
{\it Left panel:} The density map.
{\it Second Panel:} The poloidal velocity field.
{\it Third panel:} the ratio between the rotational and radial components of the velocity. Note the rapid change close to the equator ($\theta=90^\circ$) where the rotational component dominates.
{\it Right panel:} The parameter $f_k$ (Eq.~[\ref{eq:f_k}]).
\textbf{Subfigure (b):} {\it Left panel:} the $\rho^n$-weighted and surface-integrated kinetic energy, normalized by the gravitational potential.
{\it Right panel:} The components of the density-weighted and surface-integrated kinetic energy, normalized by the gravitational potential.
The components that appear in the legend but not in the figure (or in the figure partially) have lower values than the lower limit of the ordinate, and are negligible compared to the others.
\textbf{Subfigure (c):} {\it Left panel:}  Inverse of various virial factors $\mathfrak{f}_i^{-1} = 2 Q_i / |\Phi_G|$ (Eq.~[\ref{Qi}]), where $Q_i$ are the internal energy $E$, the total kinetic energy $K$, and its components $K_r$, $K_\theta$ and $K_\varphi$.
Also plotted is $M_g/M_{\rm{tot}}$, the normalized, integrated gas mass contained within radius $r$.
The dashed-gray line gives the value of $\mathfrak{f}^{-1}$ (Eq.~[\ref{eq:fancy_f}]) as a function of radius. Note that it overlaps with the blue line ($K$) as $E \ll K$ ($E$ is below the lower limit of the ordinate).
The value of $\mathfrak{f}$ is the important value for determining if the flow is virialized.
Here it is not flat and reaches values $\ll 1$, therefore the flow is not virialized.
The middle and right panels show the projected quantities $K_r$ and $K_\varphi$, respectively, observed from viewing angles $i$, measured from the pole (see legend).
}}
\label{fig:i6r}
\end{figure*}
%
%
\begin{figure*}
\textbf{\Large (a)}
\newline
\includegraphics[trim=0cm 0.17cm 0cm 0.65cm, clip=true, width=1.06\textwidth]{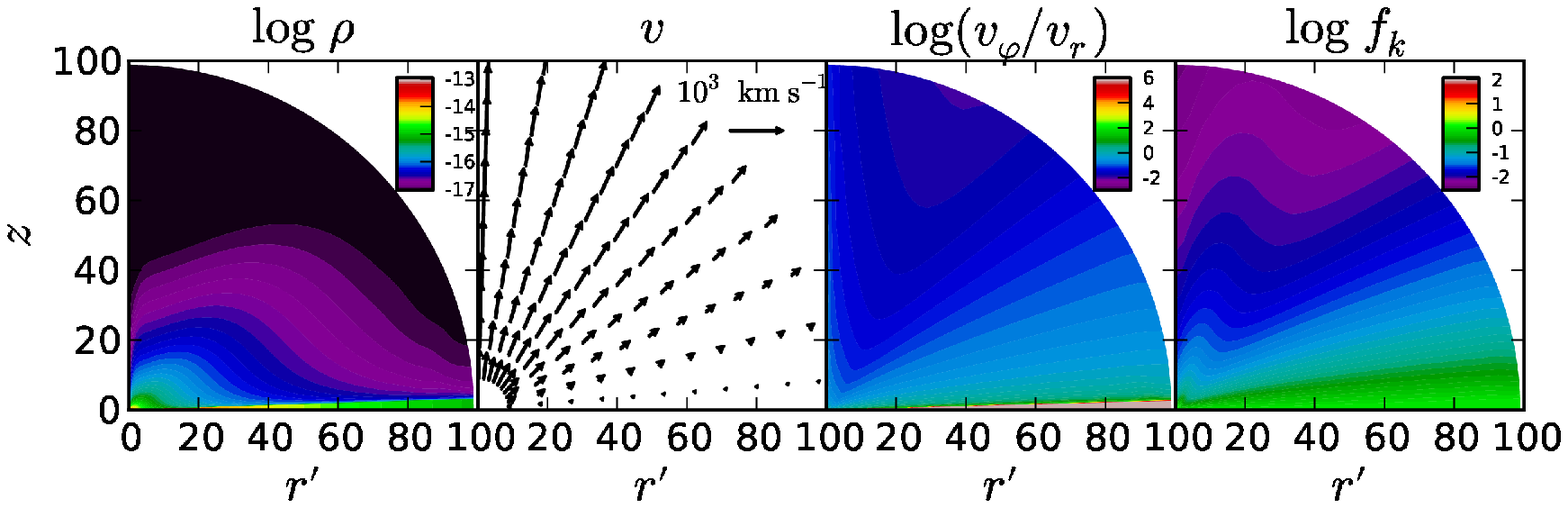}
\textbf{\Large (b)}
\newline
\includegraphics[trim=0cm 0.5cm 0cm 0.5cm, clip=true, width=1.0\textwidth]{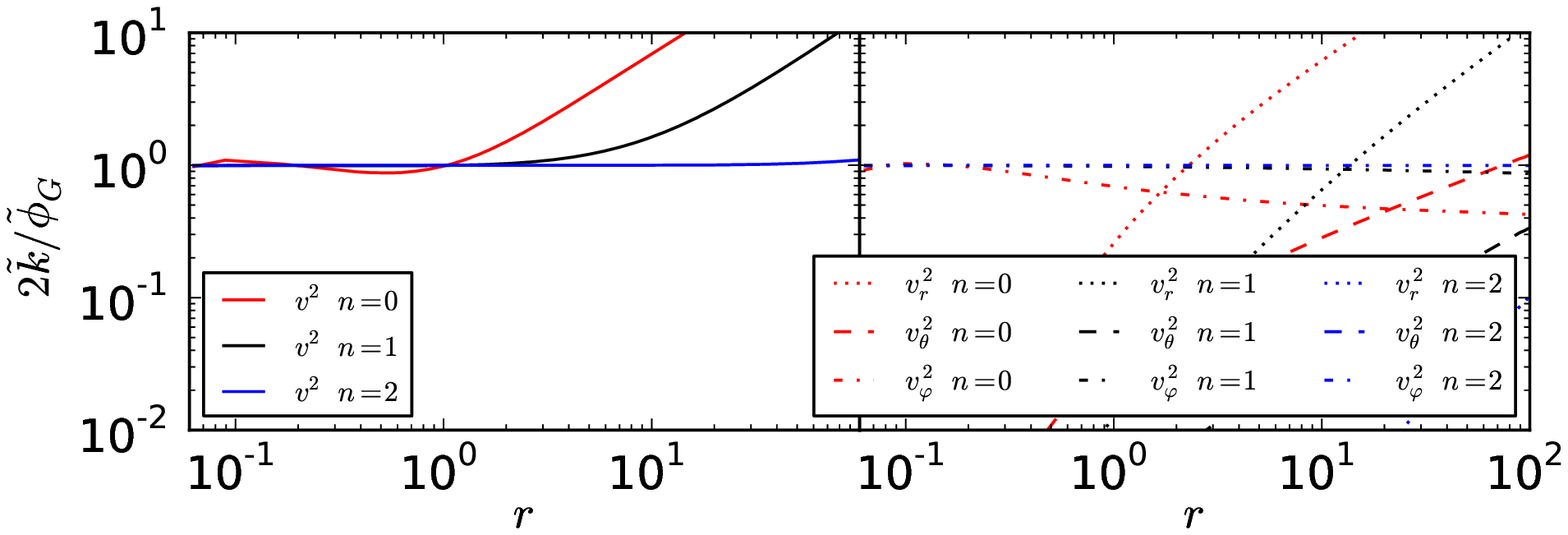}
\textbf{\Large (c)}
\newline
\includegraphics[trim=0cm 0.6cm 0cm 0.55cm, clip=true, width=1.0\textwidth]{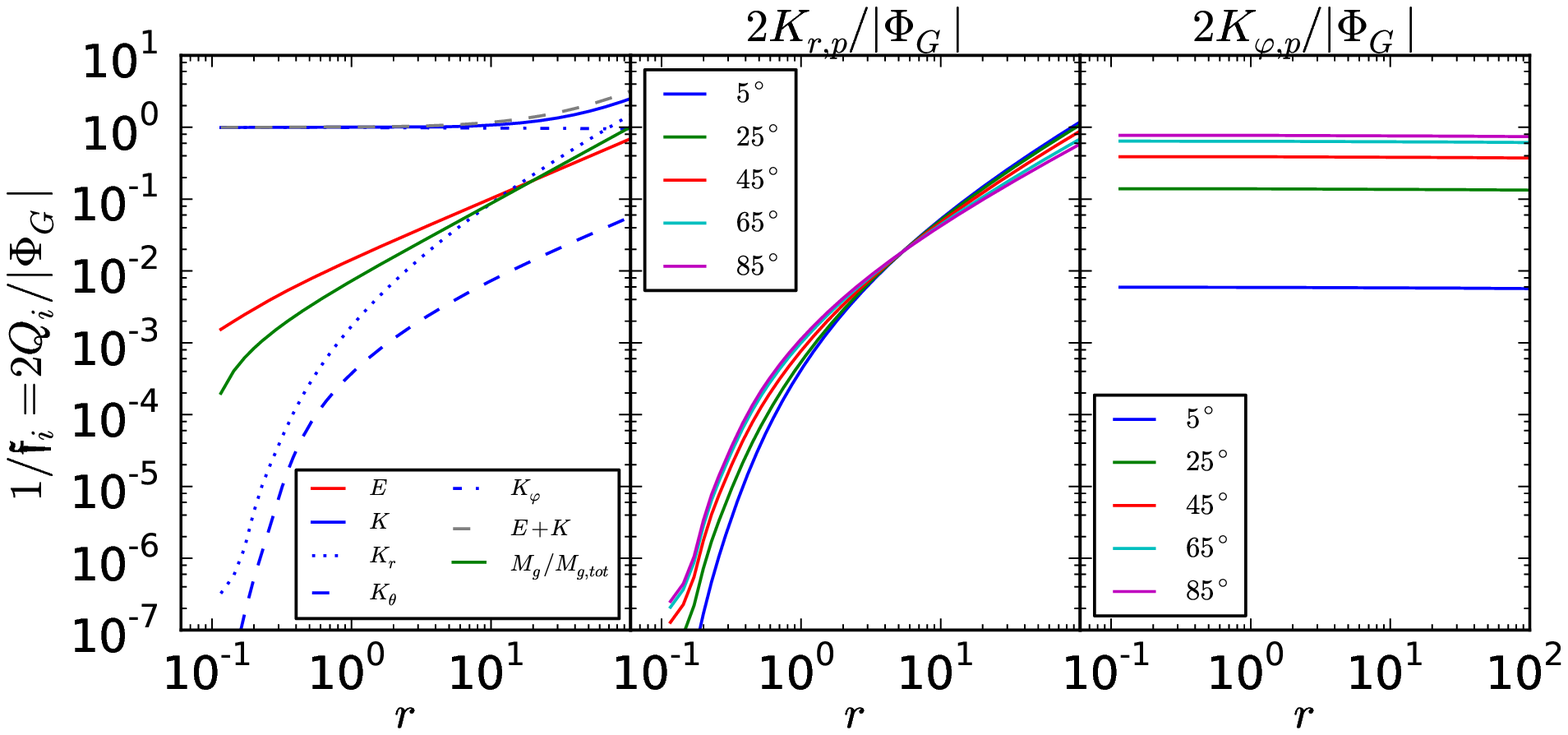}
\caption{Same as Fig.~\ref{fig:i6r}, but for Model 2. The distance for this model is in units of the inverse Compton  radius $r_{\rm{IC}}$.  Note that $r^{\prime} = r \sin \theta$. 
}
\label{fig:151dw28}
\end{figure*}
%
%
\begin{figure*}
\textbf{\Large (a)}
\newline
\includegraphics[trim=0cm 0.17cm 0cm 0.65cm, clip=true, width=1.06\textwidth]{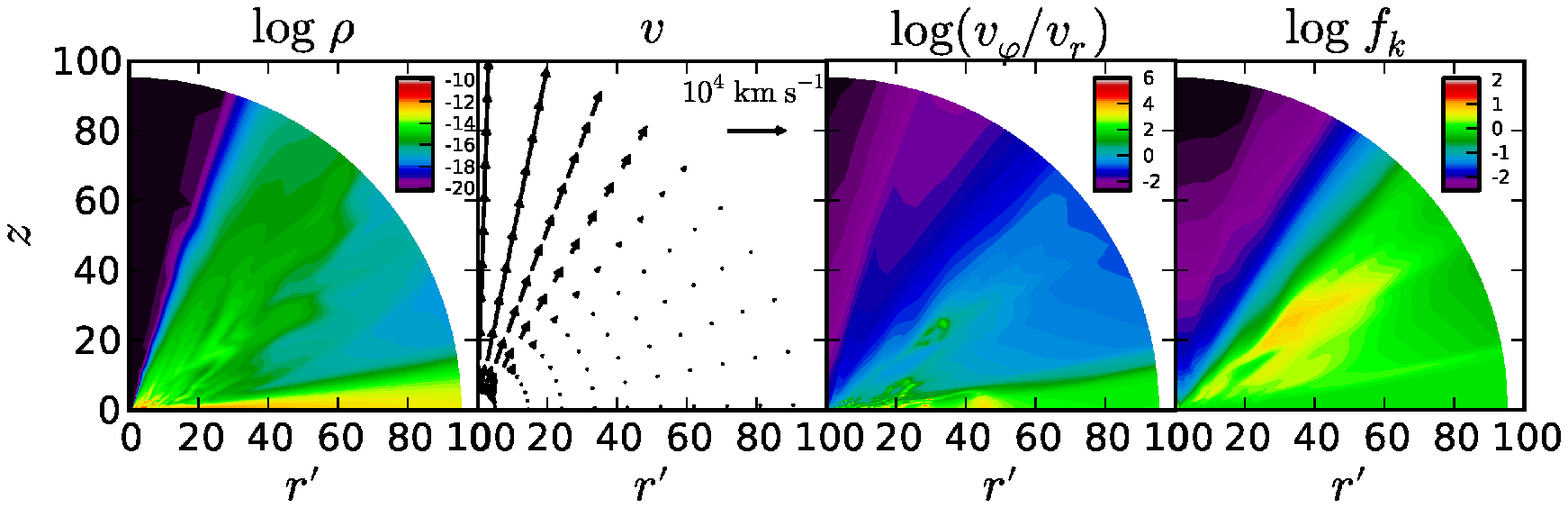}
\textbf{\Large (b)}
\newline
\includegraphics[trim=0cm 0.5cm 0cm 0.5cm, clip=true, width=1.0\textwidth]{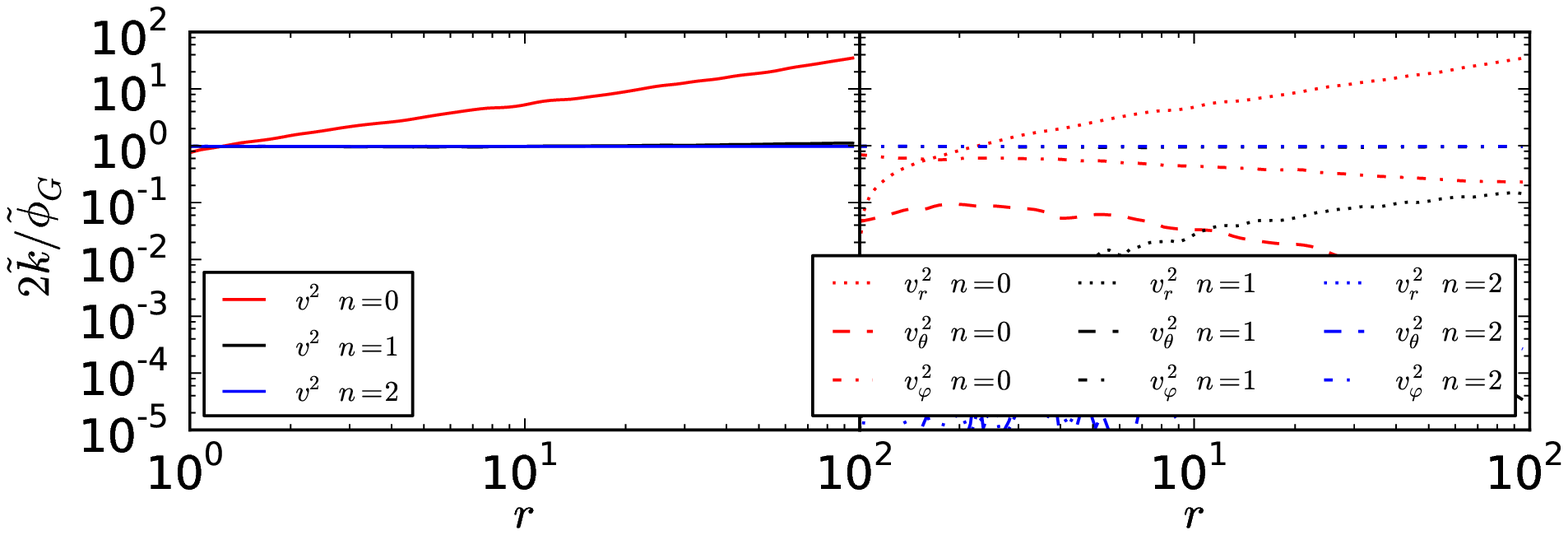}
\textbf{\Large (c)}
\newline
\includegraphics[trim=0cm 0.6cm 0cm 0.55cm, clip=true, width=1.0\textwidth]{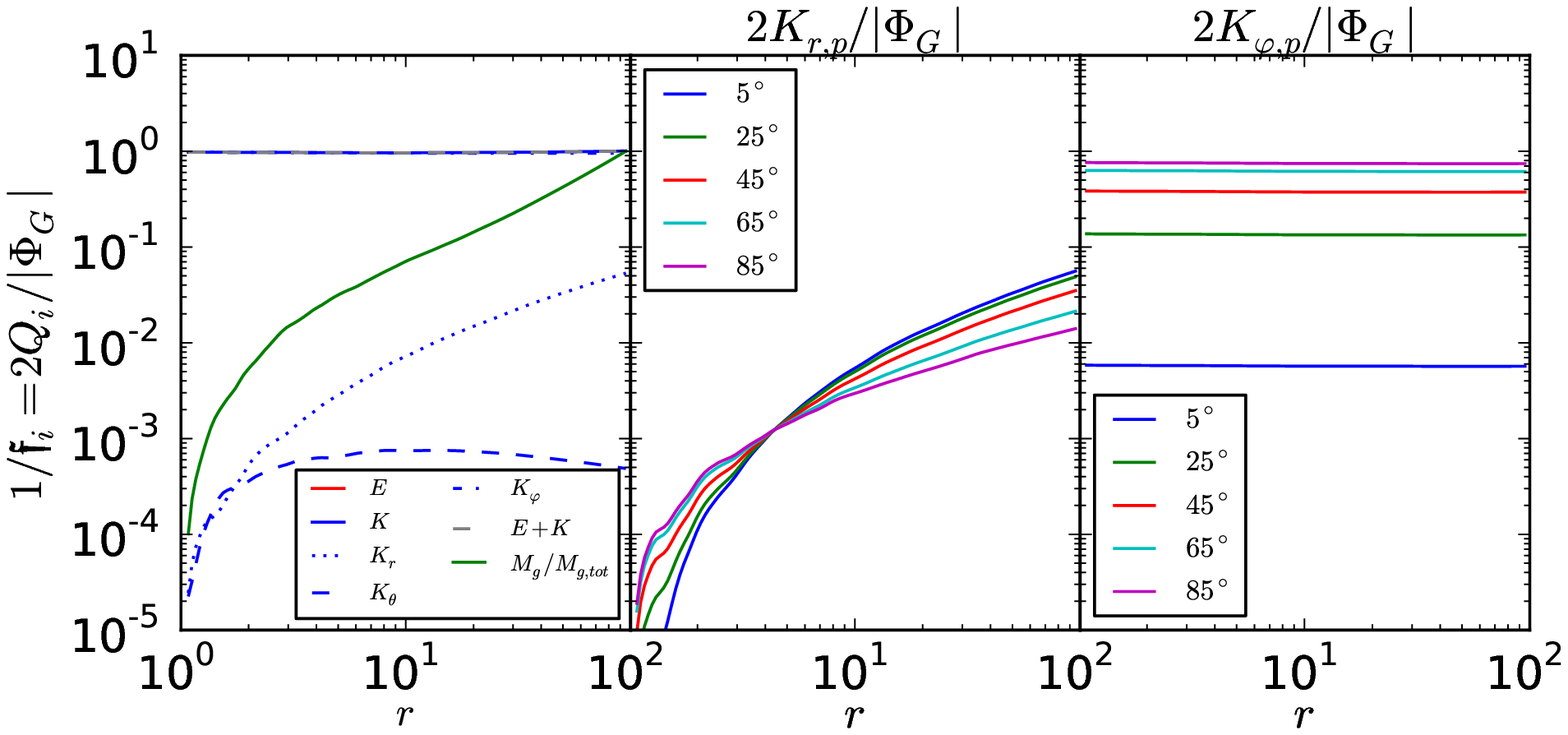}
\caption{Same as Fig.~\ref{fig:i6r}, but for Model 3. The distance for this model is in units of white dwarf radius $r_{\rm{WD}}$.
Note that, in the left panel of sub-figure (c), the dashed-gray line ($\mathfrak{f}$) overlaps with the blue line ($K$) as $E \ll K$, and with the dotted-dashed blue line $K_\varphi$, because $K_\theta,K_r \ll K_\varphi$.
The value of $\mathfrak{f}$ is flat, therefore we conclude that for this model the flow is virialized.
}
\label{fig:CV_E2}
\end{figure*}
%
%
\begin{figure*}
\textbf{\Large (a)}
\newline
\includegraphics[trim=0cm 0.17cm 0cm 0.65cm, clip=true, width=1.08\textwidth]{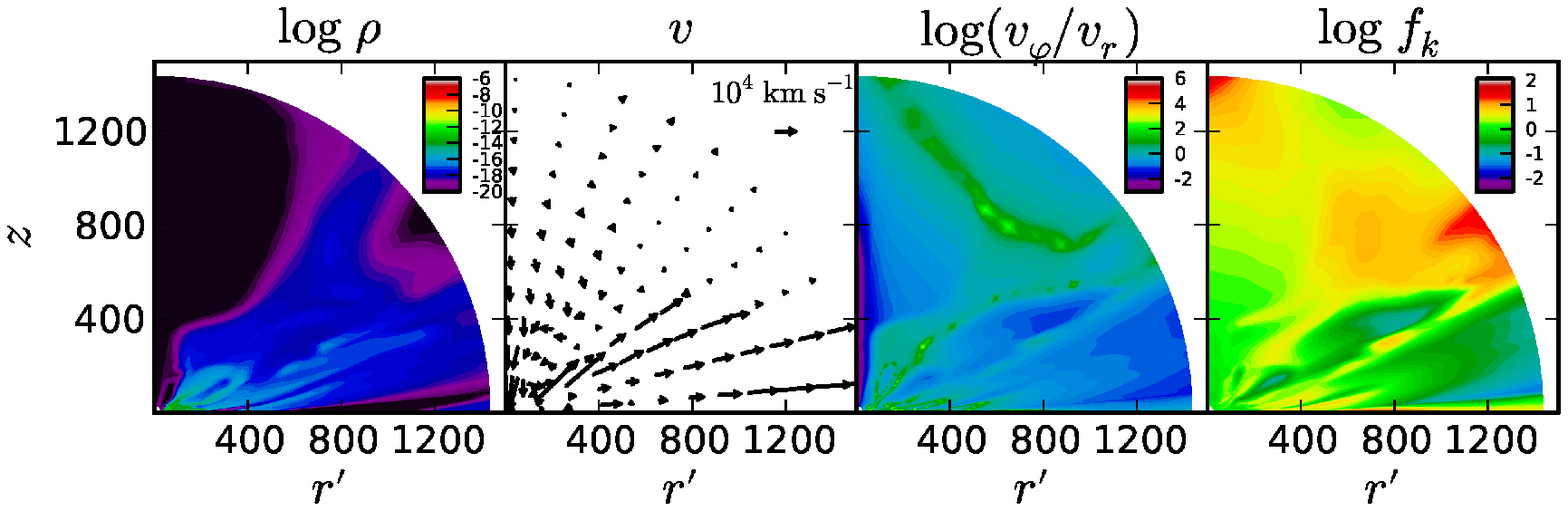}
\textbf{\Large (b)}
\newline
\includegraphics[trim=0cm 0.5cm 0cm 0.4cm, clip=true, width=1.0\textwidth]{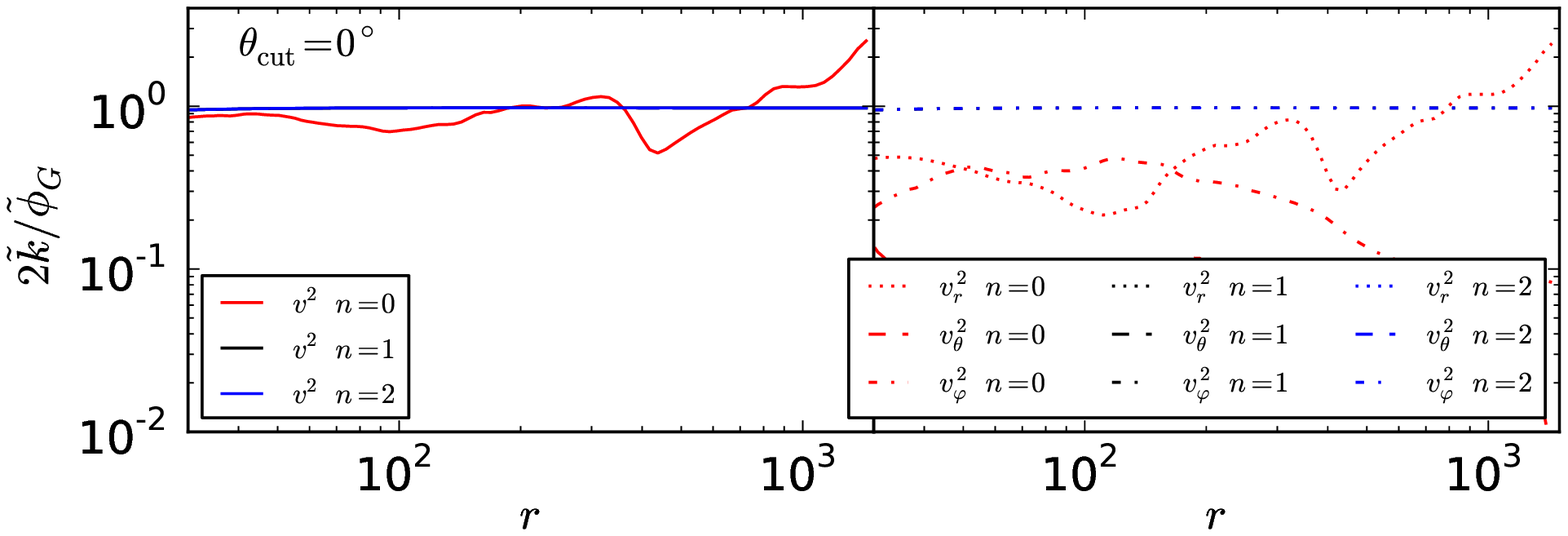}
\textbf{\Large (c)}
\newline
\includegraphics[trim=0cm 0.6cm 0cm 0.55cm, clip=true, width=1.0\textwidth]{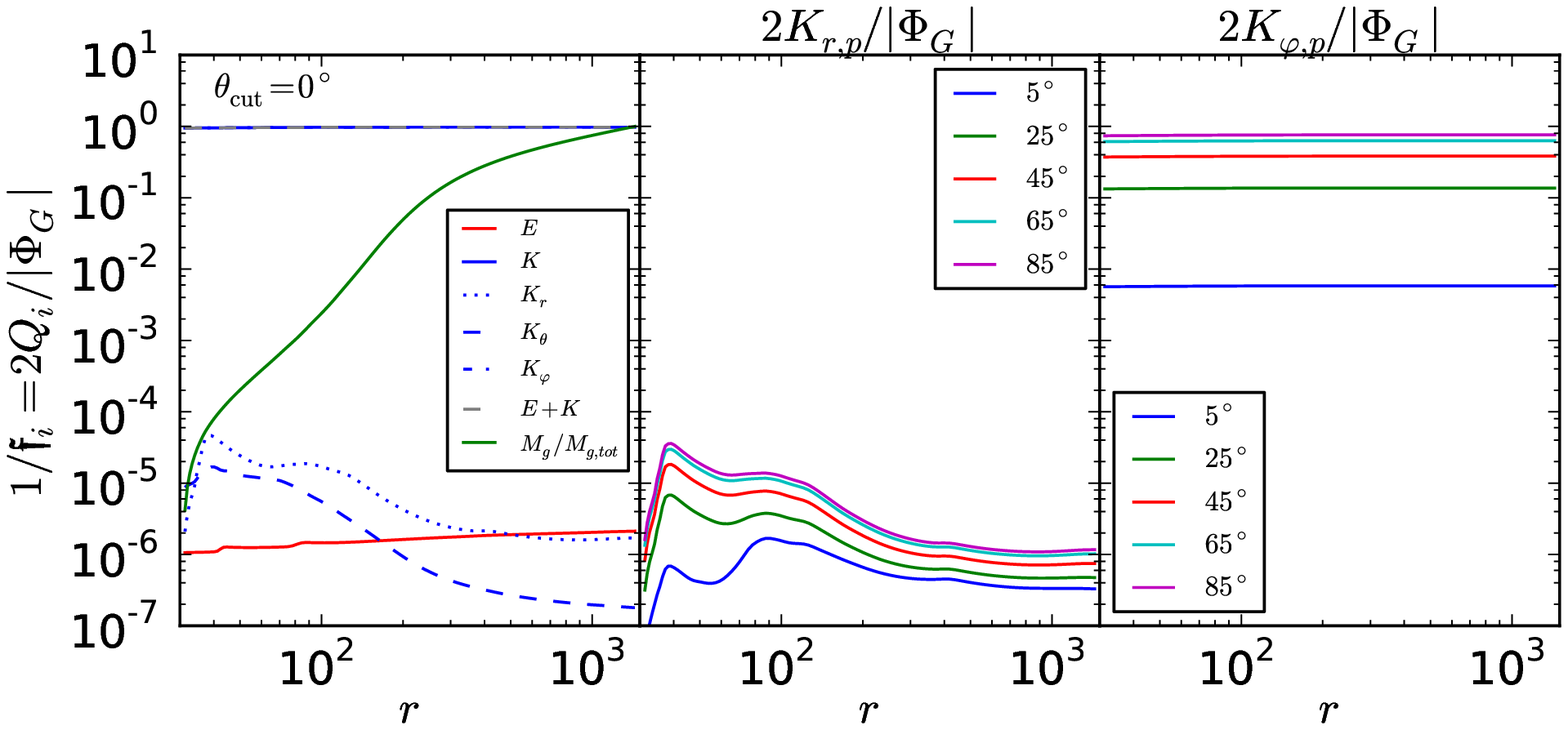}
\caption{Same as Fig.~\ref{fig:i6r}, but for Model 4.  The distance for this model is given in units of the Schwarzschild radius $r_s=2GM/c^2=3 \times 10^{13} \cm $.
Note that in the left panel, the dashed-gray line ($\mathfrak{f}$) overlaps with the blue line ($K$) because $E \ll K$ ($E$ is below the lower limit of the ordinate), and with the dotted-dashed blue line $K_\varphi$ because $K_\theta,K_r \ll K_\varphi$.
The value of $\mathfrak{f}$ is flat, therefore we conclude that for this model the flow is virialized.
}
\label{fig:KP04}
\end{figure*}

In the following, we present three sub-figures for each simulation, and each sub-figure contains a few panels. 
The {\it sub-figure (a)} presents the following flow properties:
the density map (left most panel), the poloidal velocity field (second left panel),
the comparison of radial and rotational velocities ($v_\varphi / v_r$; third panel), 
and $f_k$ (right-most panel).
Note that the abscissa of sub-figure (a) is using $r^{\prime} = r \sin \theta$. 

The {\it sub-figure (b)} presents the ratio of gravitational potential and kinetic energy $ 2 \tilde{k} / \tilde{\phi}_G $, weighted by $\rho^n$ and surface-integrated (see Eq.~[\ref{eq:tildeqi}]).  We are presenting its components in order to show the contribution of each component separately.
The left panel shows the ratios for the total kinetic energy with $n=0,1,2$, whereas the right panel shows the radial, meridional and rotational components.

Finally, the {\it sub-figure (c)} presents the volume-averaged properties 
of the flows.
The left panel presents the quantities $2Q_i / |\Phi_G| $ (Eq.~[\ref{Qi}]), which relate to each component of the inverse virial factor. 
We define the density-weighted, volume-integrated virial factor 
\begin{equation}
\mathfrak{f}=\frac{|\Phi_G|}{2(E+K)},
\label{eq:fancy_f}
\end{equation}
as an analogue of the non-density weighted integrand $f$, which is also shown in the left panel by the dashed gray line. 
The theoretical value of $\mathfrak{f}$ is $0.5$ for supersonic accretion, and $1$ for Keplerian rotation.
The same panel also shows $M_{g}/M_{\rm{tot}}$, the total gas mass within a radius $r$, normalized by the total mass  (green solid line).
In the middle and right panels, we present the projected radial and rotational components of the kinetic energy $K_{r,p}$ and $K_{\varphi,p}$ respectively, as viewed at an angle $i$, measured from the pole. 

We emphasize that the ultimate indication for a flow to be virialized is that $\mathfrak{f}$ has a flat radial profile, because these components 
(density-weighted, volume-integrated internal energy, kinetic energy 
and gravitational potential) are the quantities that enter the virial theorem.

Figure~\ref{fig:i6r} presents the results of Model 1 (the isothermal wind). 
Figure~\ref{fig:i6r}a shows that the radial component dominates over the rotational component 
everywhere except for near the equator.
The right panel shows that $f_k$ is a function of $r$, but a very weak function of $\theta$.
It is evident that at $ r \gtrsim 1$ the flow is highly non virialized for every $\theta$.

Figure~\ref{fig:i6r}b shows that the radial component of $\tilde{k}$ becomes dominant at $r \gtrsim 0.5$, independent of the value of $n$.
This transition radius is expected, because the density becomes close to spherically symmetric at greater radii, 
and thus the integration does not affect the value of $\tilde{k}$ (Eq.~[\ref{eq:tildeqi}]).

Figure~\ref{fig:i6r}c reveals four significant properties of the isothermal wind:
(i) The rotational component $K_\varphi$ is dominant up to $r \simeq 1$, and beyond this radius $K_r$ becomes dominant. 
(ii) The flow is not virialized, mainly due to the large values of $K_r$.
(iii) The projected component $K_{r,p}$ varies very little as a function of a viewing angle.  Again this is expected, because  the wind is close to spherically symmetric, and the velocity field is dominated by the radial component.
(iv) The projected component $K_{\varphi,p}$ varies as a function of a viewing angle, because $v_\varphi$ is not spherically symmetric.
We find that the kinetic properties of the flow do not scale with the gravitational potential,
independent of the density weighting and viewing angle.
This indicates that, for the isothermal wind model, the outflow would be observed as non-virialized 
in resonance and recombination lines from any viewing angle.

Figure~\ref{fig:151dw28} shows the results for Model 2 (thermal wind from an X-ray heated disk). 
The right panel of Figure~\ref{fig:151dw28}a shows that $f_k$ is a strong function of $r$ and $\theta$.
Therefore it is not straightforward to draw conclusions about the virialization of the flow using this information alone,
though the large volume with blue and purple colors ($f_k \ll 1$) hints that this is not a virialized flow.
It requires our analysis of density-weighted, surface and volume integrated quantities to verify that the flow is not virialized. 

Figure~\ref{fig:151dw28}b shows that for Model 2, the radial range where the rotational component is dominant  depends on $n$. For $n=0$ and $1$, the radial part becomes important at $r \gtrsim 1$, while for $n=2$ the rotational component dominates up to the outer computational domain of $r=100$.

From the left panel of Figure~\ref{fig:151dw28}c, we learn that the flow is virialized in the inner region of $r \lesssim 2$ (see the dashed gray line), in which point the radial component becomes more important than the rotational component. 
The rotational component originates from the Keplerian disk, and it makes the flow virialized.
One can say that the wind ``remembers'' its attachment to the central object up to that radius.
We conclude that the wind in Model~2 would be observed as non-virialized, 
because it is virialized only in the inner regions. 

Figure \ref{fig:CV_E2} present the results of Model 3 (line-driven wind).
Figure \ref{fig:CV_E2}a shows that $f_k$ varies significantly as a function of $\theta$, but is only a weak function of $r$.
More importantly, a large volume ($\theta \gtrsim 40^\circ$, for all $r$) has $f_k \sim 1$ (shown in green-to-yellow colors), which indicates that the flow is virialized.
The left-most panel shows that this region is the denser part of the flow.

Figure~\ref{fig:CV_E2}b shows that for this model, similarly to the Model 2, the radial range where the rotational component dominates depends of $n$. For $n=0$, the radial component dominates, and the kinetic component does not scale with gravity, 
while for $n=1$ and $2$ the rotational component dominates up to the outer computational domain of $r=100$.

Figure~\ref{fig:CV_E2}c shows that for this model, the flow is completely virialized up to the radial
boundary.
The right panel shows that $K_{\varphi,p}/|\Phi_G|$ is almost independent of radius. 
The actual value of $K_{\varphi,p}/|\Phi_G|$ scales as $\propto \sin^2{i}$.
We find that $K_{\varphi,p}>K_{r,p}$ for viewing angles $i \gtrsim 20^\circ$.
We conclude that for Model 3, the wind would be observed as virialized for viewing angles $i \gtrsim 20^\circ$.

Figures~\ref{fig:KP04} presents the results of Model 4 (line-driven wind with X-ray heating and cooling).
In the left panel of Figure~\ref{fig:KP04}a, we see a dense region near the center (colored in green) that corresponds to  the failed wind.
The right panel shows that $f_k \sim 1$ in most of the volume. 

Figure~\ref{fig:KP04}b shows similar results to Model 3 with regard to the dominance of the radial and rotational components.
According to Figure~\ref{fig:KP04}c, the flow is completely virialized up to the edge of the computational domain ($r=1500$). 
This translates to $\sim 4.5 \times 10^{16} \cm$ for a SMBH with $10^8 \Msun$. 
It is clearly seen that $K_\varphi$ is the dominant component.
The right panels show that $K_{\varphi,p}/|\Phi_G|$ remains flat for any line of sight,
and that $K_{\varphi,p}>K_{r,p}$ for any value of $\theta$.
The scaling $K_{\varphi,p}/|\Phi_G| \propto \sin^2{i}$ is the same 
as in Model 3.
This suggests that the system would be observed as virialized 
from any line of sight.

Analyzing Models 3 and 4, we find that, in the regions where the wind is virialized, the scaling of $\mathfrak{f}_p$ with the inclination angle is
\begin{equation}
\mathfrak{f}_p \equiv |\Phi_G|/K_p= \frac{1.32\pm0.08}{\sin^2{i}}.
\label{eq:fancyf_p}
\end{equation}

\begin{figure*}
\includegraphics[trim=0cm 0.5cm 0cm 0.5cm, clip=true, width=1.0\textwidth]{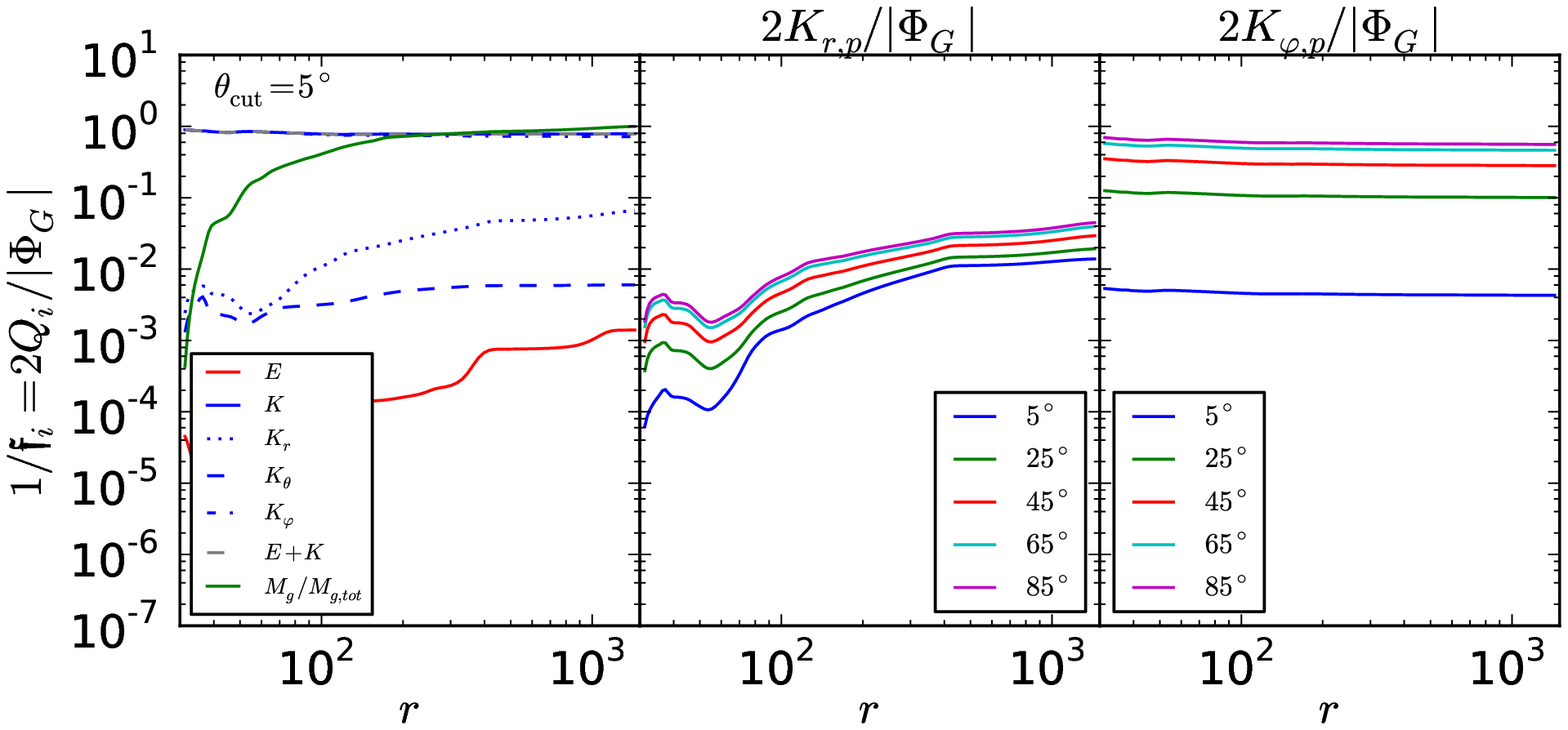}
\includegraphics[trim=0cm 0.5cm 0cm 0.5cm, clip=true, width=1.0\textwidth]{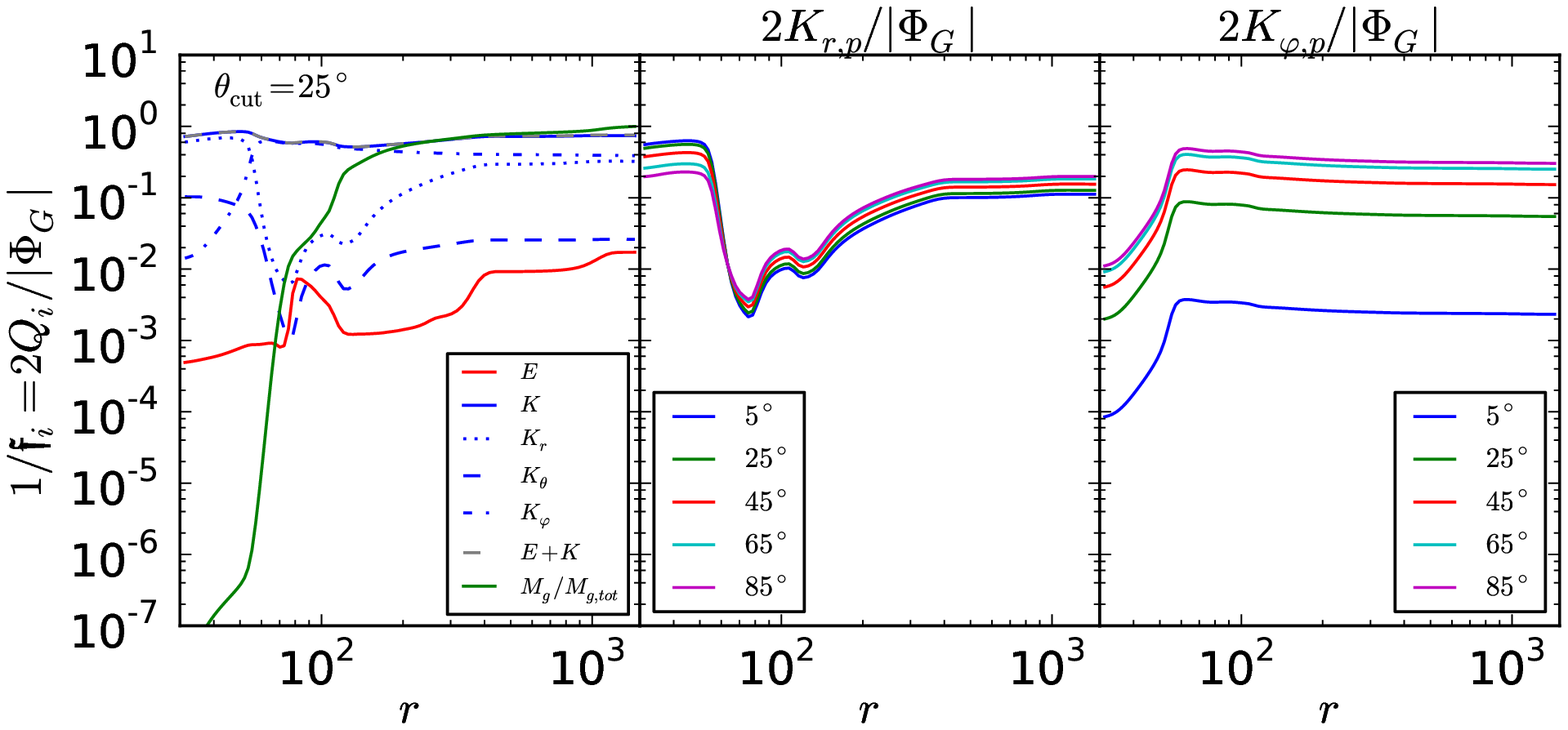}
\caption{
Same as Fig.~\ref{fig:KP04}, but with $\theta_{\rm{cut}}=5^\circ$ (top row),
and $\theta_{\rm{cut}}=25^\circ$ (bottom row). We omit the disk in integration to emphasize the wind properties.
We find that most of the mass is at low radii, and the wind is still virialized.}
\label{fig:KP04_Bernoulli_components_radial_profile_multiplot_5deg}
\end{figure*}

Another issue in testing the virialization of winds is the separation of winds from the dense disk. 
As a test, we repeat the same calculation after excluding the equatorial region with $\pi/2+\theta_{\rm{cut}}<\theta<\pi/2+\theta_{\rm{cut}}$ as follows: 
\begin{equation}
\begin{split}
Q_i &= \int\limits_{\varphi=0}^{2\pi}\ \int\limits_{\theta=0}^{\pi/2-\theta_{\rm{cut}}}\ \int\limits_{r=0}^r \ q_i \rho r^2 \sin\theta \,dr\,d\theta\,d\varphi \\
    &+\int\limits_{\varphi=0}^{2\pi}\ \int\limits_{\theta=\pi/2+\theta_{\rm{cut}}}^{\pi}\ \int\limits_{r=0}^r \ q_i \rho r^2 \sin\theta \,dr\,d\theta\,d\varphi.
\end{split}
\label{Qi_cut}
\end{equation}
We performed this test for Models 3 and 4, and found that the winds are virialized in both of cases. 
Here we show only the results for Model 4 as an example of this test.
Figure~\ref{fig:KP04_Bernoulli_components_radial_profile_multiplot_5deg} shows the result of Model 4 with $\theta_{\rm{cut}}=5^\circ$ (top row)
and $\theta_{\rm{cut}}=25^\circ$ (bottom row).
This emphasizes the significance of high-density wind near the center, 
which is shown in the left panel of Figure~\ref{fig:KP04}a in green. 
When the integrated quantities are weighted by density, it is seen that the wind itself is virialized, because $\mathfrak{f} \simeq 1$ even for $\theta_{\rm{cut}}=25^\circ$.

\section{Discussion: Unvirialization of an Outflow}
\label{sec:Discussion}

\begin{figure}
\includegraphics[width=1.0\columnwidth]{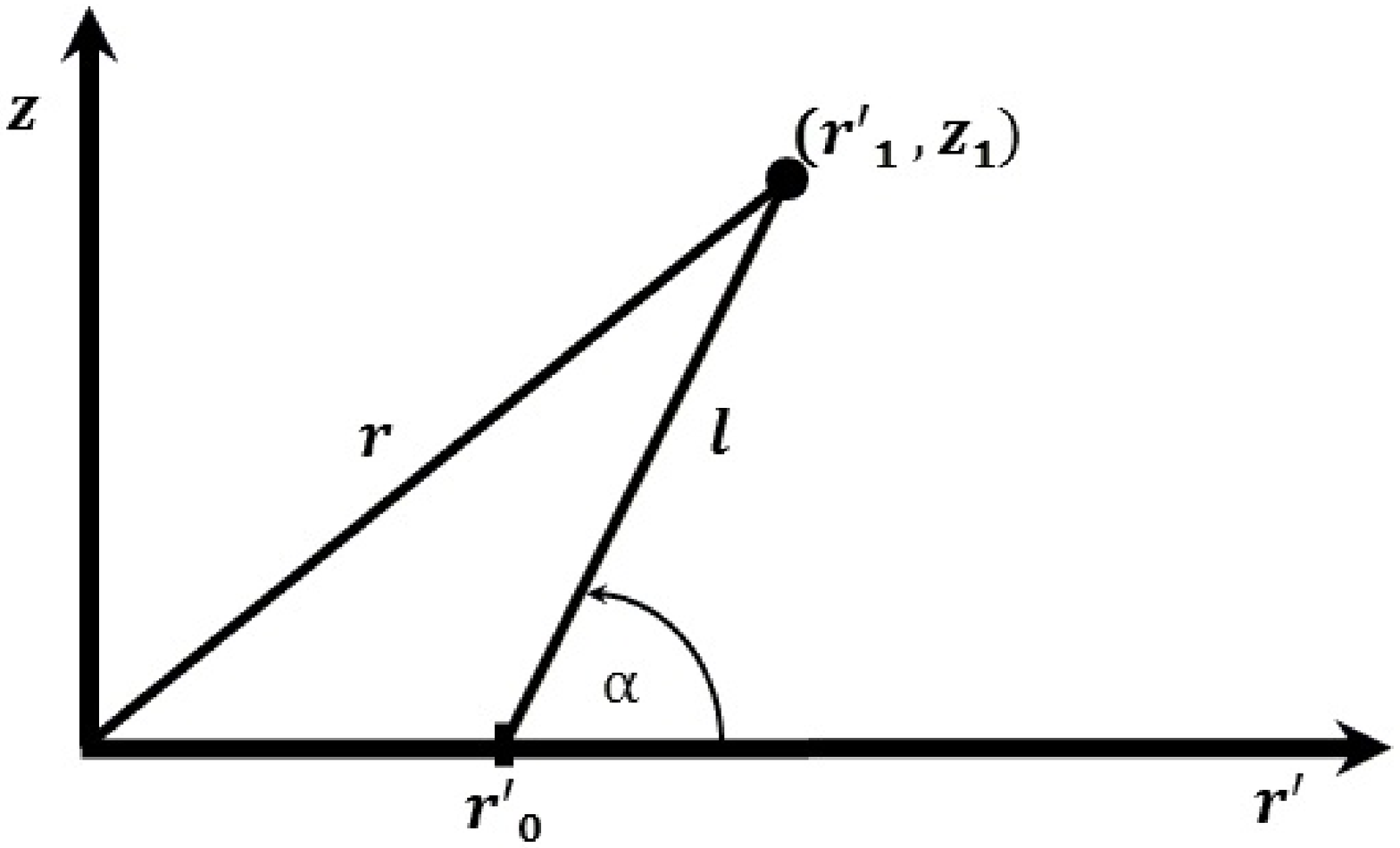}
\includegraphics[width=1.0\columnwidth]{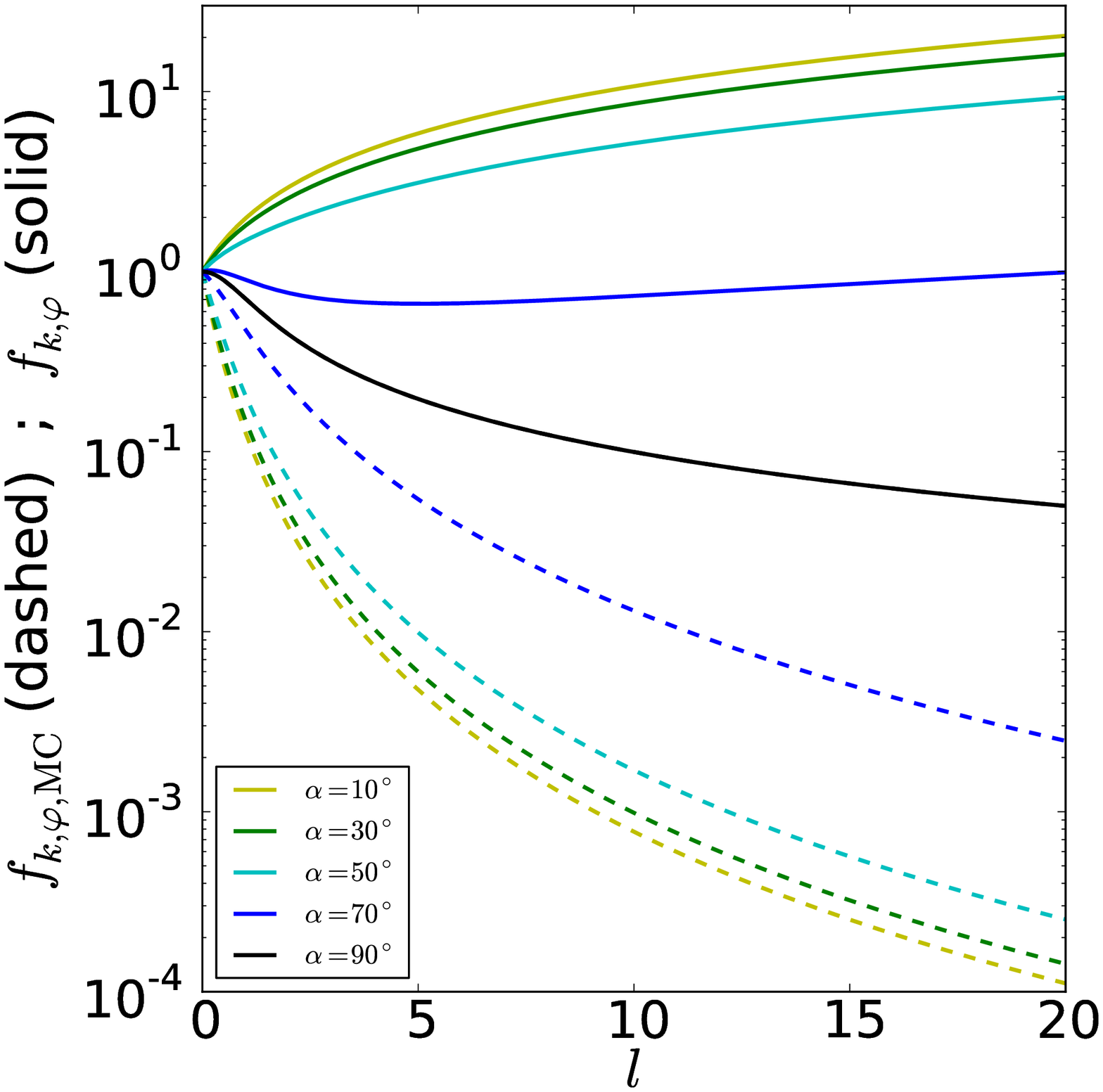}
\caption{
{\it Top panel:} Geometry of the calculation given in Section~\ref{sec:Discussion}, and relevant parameters.
{\it Bottom panel:} The rotational component $f_{k,\varphi}$ of the virial factor as a function of $l$ for the specific angular momentum conserving wind  (Eq.~[\ref{eq:vphi_over_vv}]; solid lines), and for the angular velocity conserving wind ($f_{k,\varphi,{\rm{MC}}}$, Eq.~[\ref{eq:vphi_over_vv_MC}];
dashed lines), for different values of $\alpha$ (see legend).
The wind launching point is $r'_0=1$. Note that for $\alpha=90^\circ$ the lines overlap for the two cases.
A Keplerian flow becomes unvirialized slowly when the specific angular momentum is conserved,
but much faster when the angular velocity is conserved.
}
\label{fig:Keplerian_velocity_decrease}
\end{figure}

To illustrate how a Keplerian flow becomes unvirialized due to an outflow, we use the following analytic calculation.
We consider a gas element outflowing from a point $(r'=r'_0,z=0)$ on the equator where the Keplerian velocity is $v_k=(GM/r'_0)^{1/2}$.
A new position of the element $(r'_1,z_1)$ can be expressed by the distance $l$ from the original location and the inclination angle $\alpha$ measured from the equator (see top panel of Figure \ref{fig:Keplerian_velocity_decrease}).
As a result of specific angular momentum conservation, the rotational velocity of the gas at the new location would be $v_\varphi=(r'_0/r'_1)v_k$.
We can calculate $f_{k,\varphi} = |\phi_G| / v_\varphi^2$, the rotational component of $f_k$, as a function of $l$ and $\alpha$:
\begin{equation}
f_{k,\varphi}=\frac{(r'_0+l\cos\alpha)^2}{ r'_0 [(r'_0+l\cos\alpha)^2 +(l\sin\alpha)^2]^{1/2} } .
\label{eq:vphi_over_vv}
\end{equation}
The solid lines in the bottom panel of Figure~\ref{fig:Keplerian_velocity_decrease} show $f_{k,\varphi}$ as a function of $l$ for different values of $\alpha$, for a launching point of $r'_0=1$.
One can see that generally $f_{k,\varphi}$ is a weak function of $l$.
For example, at $l=10$, $f_{k,\varphi}$ varies by a factor 
of $\sim 1.5$ at most for $\alpha=70^\circ$,
and increases by a factor of $\sim 11$ for $\alpha=10^\circ$, 

We also check another type of flow, where the conserved quantity is not the angular momentum but instead the angular velocity, as in the so-called magneto-centrifugal winds (\citealt{BlandfordPayne1982}).
In this case, $v_\varphi=(r'_1/r'_0)v_k$. As an analogue of $f_{k,\varphi}$, we calculate
\begin{equation}
f_{k,\varphi,{\rm{MC}}}=\frac{r^{\prime 3}_0}{ ( r'_0+l\cos\alpha )^2 [(r'_0+l\cos\alpha)^2 +(l\sin\alpha)^2]^{1/2} }.
\label{eq:vphi_over_vv_MC}
\end{equation}
The dashed lines in Figure \ref{fig:Keplerian_velocity_decrease} present $f_{k,\varphi,{\rm{MC}}}$ as a function of $l$ for $r'_0=1$ and different values of $\alpha$.
Comparing to the hydrodynamical case,
at $l=10$,
$f_{k,\varphi,{\rm{MC}}}$ decreases by a factor of $\sim 77$
for $\alpha=70^\circ$, ,  
while for $\alpha=10^\circ$ it decreases and by a factor of $\sim 1300$!

We find then that $f_{k,\varphi,{\rm{MC}}}$ is a much stronger function of 
$l$ than $f_{k,\varphi}$.
This indicates that an angular velocity conserving flow will become unvirialized much faster than an angular momentum conserving Keplerian flow.

We find that line-driven winds (Models 3 and 4) stay virialized over a long distance from the launching point, 
because they have an extended base of accelerating region in the poloidal wind.
Any quantity weighted by density will be dominated by the contribution from this dense base.
In Model 4, the inner region of the wind is particularly dense as it includes the failed wind (\citealt{Proga2000}; \citealt{ProgaKallman2004}).

Our analysis suggest that the flow will remain virialized under 
the following general conditions:
\begin{itemize}
\item The acceleration is vertical rather than radial
(i.e., large $\alpha$: see  Figure~\ref{fig:Keplerian_velocity_decrease}).
\item If $\alpha$ happens to be small, 
the flow better conserves the specific angular momentum rather 
than the angular velocity.
\item The wind acceleration in the poloidal direction should be slow so that the wind base will be dense.
\end{itemize}

As expected, there are cases where winds are not virialized (e.g., Parker wind in Appendix A).
However, if the wind is launched from a virialized system such as a Keplerian disk, the wind can appear as virialized to
a distance greater than $10$ times the launching radius (Figure ~\ref{fig:Keplerian_velocity_decrease}).
Therefore, our results support the assumption that the BLRs are virialized up to a large radius.

\section{Summary}
\label{sec:Summary}

The assumption that the BLRs are virialized is commonly used to determine the mass of SMBHs.
While it is widely assumed that accreting gas is virialized, winds are
often considered to be 
non-virialized as the non-gravitational forces break their connection to the central SMBH.
In the present work, we analyze simulations of winds for four 
different cases, 
and show that the flow in the line-driven disk wind ``remembers'' its
origin  -- a Keplerian disk -- 
for a relatively long distance, and as such, can be considered to be virialized. 

We also performed the same analysis for accretion/inflow simulations for 
all the models in \cite{Proga2007} and \cite{KurosawaProga2009},
covering a large range of parameter space (Eddington ratio, density,
temperature, X-ray 
fraction and more), including models where thermal instability created dense cool clumps \citep[cf.,][]{Barai2011, Barai2012, Moscibrodzka2013}.  
Generally, all inflows were found to be virialized
because they are supersonic. 
As these results are similar to the above four simulations, we do not show them here.

We find that if the emission lines are generated close to the center, they would be observed as virialized up to a large radius.
For a typical AGN with a SMBH of $10^8 \Msun$, the flow would be virialized up to $ \gtrsim 10^3 r_s$, where BLRs would be included.
The projected quantities (the left panel of Figures~\ref{fig:i6r}c, \ref{fig:151dw28}c, \ref{fig:CV_E2}c and \ref{fig:KP04}c)
show that $K_{\varphi,p}$ dominates over $K_{r,p}$ even at small inclination 
angles, up to $\sim 20^\circ$ (for higher $i$
$K_{r,p}$ becomes dominant).
The implication of this is that the observed projected value of $\mathfrak{f}$ will scale as $\mathfrak{f} \propto 1/\sin^2{i}$ for $i \gtrsim 20^\circ$.
As $\mathfrak{f}_p$ and $f$ relate to each other by density weighting and integration over volume,  this implies that $f \propto 1/\sin^2{i}$.

The average factor $\langle f \rangle$ discussed in the literature ranges from $\langle f \rangle = 1$ (\citealt{McLureDunlop2004}), and raging up to $\langle f \rangle = 4-6$ \citep{Onken2004, Woo2010, Grier2013}. 
Our Equation~(\ref{eq:fancyf_p}) is consistent with the result of \cite{Onken2004}, who derived $f=2 \ln{2} / \sin^2{i}$, assuming a thin ring in Keplerian rotation.
The dependency of $f \propto 1/ \sin^2{i}$ is indeed expected, as the flow in our simulations can be thought of as a collection of rings in Keplerian rotation.
The numerical factor $2 \ln{2}$ by \cite{Onken2004} comes from relating the rotational velocity to the line velocity dispersion,
while our numerical factor $1.32\pm0.08$ comes from density weighting and integration.
Therefore we regard the numerical consistency as coincidental.
We shall return to this point in a future paper where synthetic line profiles will be computed.

We note that other studies give different results for $f(i)$. For example, \cite{Decarli2008} suggested $f=(2c_1 \sin{i} +2c_2/\sqrt{3})^{-1/2}$, where the parameters $c_1$ and $c_2$ quantify the importance of the disk and isotropic components of the BLR, respectively.  
For a thin disk $c_1 \rightarrow 1$ and $c_2 \rightarrow 0$, then $f \approx 1/\sqrt{2 \sin{i}}$.

Theoretically, the uncertainty in the value of $f$ is the source of error in the determination of SMBH masses. 
However, the theoretical value of $f$ is not necessarily identical to 
the $f$-factor that is used in observational estimates.
The latter depends on the particular way the line width is measured (e.g., whether FWHM or $\sigma$ is used to determine the width,
or how the line-width is determined for a profile with multiple peaks).

Furthermore, the BLR radius obtained by reverberation mapping is a weighted average, 
estimated from the cross-correlation lag between the changes of continuum and line response.
In other words, the line width and the lag do not necessarily correspond to the same physical location (i.e., the
line width does not necessarily give us the circular velocity at the radius obtained from the cross-correlation lag).
We expect that there would be a simple relation between the theoretical value of $f$ and the observed one, 
however, estimating this relationship accurately is beyond the scope of this paper.

We find that the line-driven disk wind models discussed here (Models 3 and 4) are virialized to a relatively large extent,
although certainly there are classes of outflows that are not virialized (such as the Parker wind and Models 1 and 2).
Our results support the notion that the BLR can be virialized even if it contains an outflow, provided that the outflow's structure is
similar to that of a dense, slowly accelerating line-driven wind from a Keplerian disk.

This, in turn, gives additional support for a two component model (disk+line-driven disk wind) of
the BLR in general and not just for the outflows in particular
\citep[see][who arrived at a similar conclusion by considering only the disk and the wind base not the whole wind]{ChiangMurray1996}.
Clearly it is necessary to continue to test this model.
Our conclusions here motivate calculations of the reverberation response of a disk wind even if it is complex and time-dependent as in Model 4.

\section*{Acknowledgments}
We thank Alexei Baskin, Shai Kaspi, Julian Krolik, Ari Laor, Anna Pancoast, Stephen Rafter, Noam Soker, Timothy Waters and an anonymous referee for helpful comments.
We acknowledge support provided by NASA through grants HST-AR-12835 and HST-AR-12150.01-A.
from the Space Telescope Science Institute, which is operated by the
Association of Universities for Research in Astronomy, Inc., under NASA contract~NAS5-26555.
This work was also supported in part by the NSF grant AST-0807491.
Research by A.J.B. is supported by NSF grant AST-1108835.  

\appendix

\section{Bondi Accretion And Parker Wind Virial Quantities}

\begin{figure*}
\includegraphics[width=1.0\textwidth]{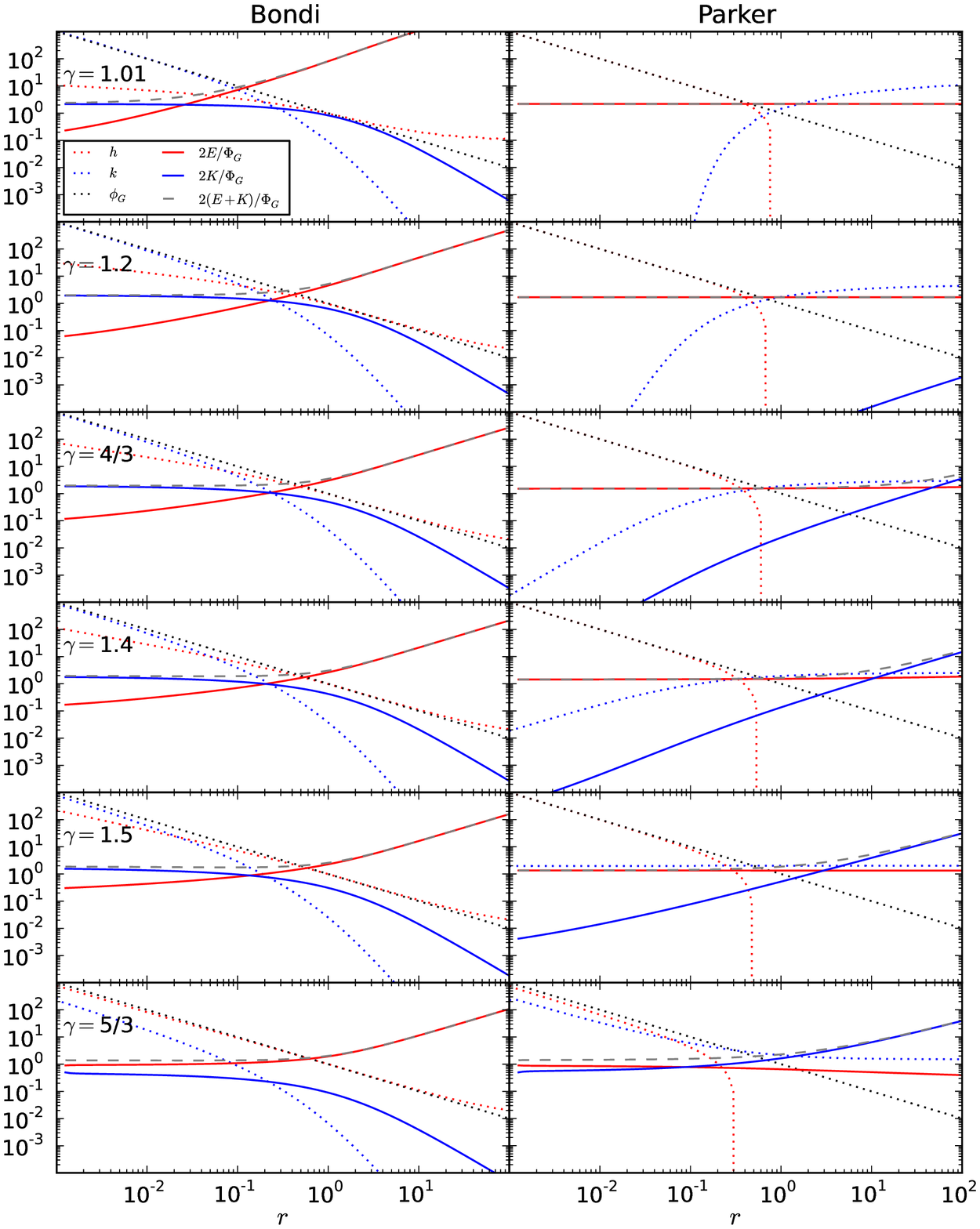}
\caption{
{\it Left panels:} Dotted lines show the specific kinetic energy $k$, specific enthalpy $h$, and gravitational potential $\phi_G$, calculated by solving the 1D Bondi accretion problem.
Solid lines are the density-weighted, volume-integrated quantities $Q_i$ of the virial theorem: kinetic energy ($K$), enthalpy $E$, and gravitational potential $\Phi_G$.  Plotted are the ratios $\mathfrak{f}_i^{-1} = 2Q_i/|\Phi_G|$.
{\it Right panels:} Same as the left panels, but for the Parker wind solution.
}
\label{fig:Bondi_Parker}
\end{figure*}

We demonstrate our analysis method on both Bondi accretion \citep{Bondi1952} and Parker winds (\citealt{Parker1965}; see also \citealt{WatersProga2012} and references therein).
The classic \cite{Bondi1952} accretion scenario assumes a steady state where gas is in a radial flow with two forces acting on it: gravity and gas pressure.
The flow is into a central point mass $M$, which is the only source of gravity.
A steady accretion flow satisfies the Bernoulli equation
\begin{equation}
\frac{v^2}{2} + \int_{P_\infty}^{P} \frac{dP}{\rho} - \frac{GM}{r} = 0, 
\label{eq:Bernoulli_pure}
\end{equation}
where $v$ is the flow speed, $G$ is the gravitational constant, and $r$ is the radial distance from central object with mass $M$.
The pressure $P$ and the gas density $\rho$ are assumed to be uniform at infinity with values $P_\infty$ and $\rho_\infty$, respectively,
and satisfy the polytropic equation of state
$P/P_\infty= (\rho/\rho_\infty)^{\gamma}$, 
where $\gamma$ is the adiabatic index.
The speed of sound at infinity is $c_{s,\infty}=\sqrt{\gamma P_\infty/\rho_\infty}$.
The Bernoulli equation with the above equation of state is then
\begin{equation}
\frac{v^2}{2} + \left(\frac{\gamma}{\gamma-1}\right) \frac{P_\infty}{\rho_\infty} \left[\left(\frac{\rho}{\rho_\infty}\right)^{\gamma-1} -1 \right] - \frac{GM}{r} = 0.
\label{eq:Bernoulli}
\end{equation}
The continuity equation gives the accretion rate $\dot{M} = \int r^2 \rho v \, d\Omega$. 
Under the spherical symmetry, $\dot{M}=4 \pi r^2 \rho v$.
The solution gives the
the Bondi accretion rate
\begin{equation}
\dot{M}_{\rm B} =  \frac{\lambda 4\pi G^2 M^2 \rho_\infty}{c_{s,\infty}^3},
\label{eq:MdotBondi}
\end{equation}
where the value of the constant $\lambda$ determines the solution.
A characteristic radius is also defined -- the Bondi radius $R_B=GM/c_{s,\infty}^2$ -- the distance from $M$ where the forces are balanced.

Depending on the value of $\lambda$, a range of solutions are possible (Fig.~2 in \citealt{Bondi1952}); 
the one relevant for astrophysical accretion is    
the so-called critical solution with the value of 
\begin{equation} 
\lambda_c = 
\left( \frac{1}{2} \right)^{ \frac{ \left( \gamma + 1 \right) } { 2 \left( \gamma - 1 \right) } }  
\left( \frac{5 - 3 \gamma}{4} \right)^{ \frac{ \left( 3 \gamma - 5 \right) } { 2 \left( \gamma - 1 \right) } } . 
\end{equation} 
For the critical solution, gas is subsonic in the outer parts, traverses a sonic point, 
and accretes onto the central object at a supersonic velocity.
The second solution with the same value $\lambda_c$ resembles a Parker wind.
A supersonic inflow exists for $r<R_s/ R_B = (5 - 3 \gamma)/4$ if $\gamma<5/3$, and there the gas is free falling.
For a supersonic inflow, an analytic estimate is possible in the limit  of $r \ll 1$.

We solve the classical 1D analytic solutions of the Bondi problem to find the density,
velocity (and consequently the pressure and temperature) as a function of radius.
We calculate the kinetic energy $k=v^2/2$, the gravitational potential due to the central mass $\phi_G=-GM/r$, and the specific enthalpy $h=\gamma e = (\gamma/\gamma-1) P/\rho$, where $e$ is the specific internal energy. 
We add a constant $-GM/[R_B(\gamma-1)]$ to the specific enthalpy to satisfy Equation~\ref{eq:Bernoulli}.
The above quantities are then presented in dimensionless units, taking $GM=1$ (Fig. \ref{fig:Bondi_Parker}).

We calculate the density-weighted, volume-integrated quantities of the virial theorem as 
\begin{equation}
Q_i = 4\pi \int\limits_{r=0}^r \ q_i \rho r^2 \,dr, 
\label{Qi_BondiParker}
\end{equation}
where $q_i = (k, e, \phi_G)$ and $Q_i = (K, E, \Phi_G)$, respectively.

We apply the above calculation for Bondi accretion and Parker winds for different values of $\gamma$ (Fig. \ref{fig:Bondi_Parker}).
Taking the free-fall density and velocity, and substituting them into Eq.~\ref{Qi_BondiParker}, we find that the free-fall limit gives $|\Phi_G|/ 2K = 0.5$.  This is an expected result as a radial virialized inflow gives $f=0.5$, 
and as long as both sides of the equation are weighted by the same density profile the result should remain unchanged.


{}

\end{document}